\documentclass[twocolumn,twoside,slac_two]{revtex4}
\usepackage{graphicx}
\usepackage{fancyhdr}
\begin{document}

\title{Observations of Soft Gamma Ray Sources $>100$ keV Using Earth Occultation with GBM}

%

\author{G.L. Case, M.L. Cherry, J. Rodi }
\affiliation{Dept. of Physics \& Astronomy, Louisiana State Univ., Baton Rouge, LA 70803, USA}
\author{A. Camero-Arranz}
\affiliation{Fundaci\'{o}n Espa\~{n}ola de Ciencia y Tecnolog\'{i}a (MICINN), C/Rosario Pino,14-16, 28020-Madrid, Spain}
\author{E. Beklen}
\affiliation{Middle East Technical University (METU), 06531, Ankara, Turkey}
\author{C. A. Wilson-Hodge}
\affiliation{NASA Marshall Space Flight Center, Huntsville, AL 35812}
\author{P. Jenke}
\affiliation{NASA Postdoctoral Program Fellow, NASA Marshall Space Flight Center, Huntsville, AL 35812}
\author{P.N. Bhat, M.S. Briggs, V. Chaplin, V. Connaughton, R. Preece}
\affiliation{University of Alabama in Huntsville, Huntsville, AL 35899}
\author{M.H. Finger}
\affiliation{USRA, National Space Science and Technology Center, Huntsville, AL 35899}

\begin{abstract}
The NaI and BGO detectors on the Gamma ray Burst Monitor (GBM) on Fermi are now being used for long term monitoring of the hard X-ray/low 
energy gamma ray sky. Using the Earth occultation technique demonstrated previously by the BATSE instrument on the Compton Gamma Ray Observatory, 
GBM produces multiband light curves and spectra for known sources and transient outbursts in the 8 keV - 1 MeV band with its NaI detectors 
and up to 40 MeV with its BGO. Coverage of the entire sky is obtained every two orbits, with sensitivity exceeding that of BATSE at energies below $\sim25$ keV 
and above $\sim1.5$ MeV. We describe the technique and present preliminary results after the first $\sim17$ months of observations at energies above 100 keV. 
Seven sources are detected: the Crab, Cyg X-1, Swift J1753.5-0127, 1E 1740-29, Cen A, GRS 1915+105, and the transient source XTE J1752-223.

\end{abstract}

\maketitle

\thispagestyle{fancy}


\section{INTRODUCTION}
The Gamma ray Burst Monitor (GBM) on Fermi is currently the only instrument in orbit providing nearly continuous full sky coverage in the hard X-ray/low energy 
gamma ray energy range. The Earth occultation technique, used very successfully on BATSE, has been adapted to GBM. An initial catalog of 64 sources is currently 
being monitored and continuously augmented. At energies above 100 keV, six steady sources (the Crab, Cyg X-1, Swift J1753.5-0127, 1E 1740-29, Cen A, 
GRS 1915+105) and one transient source (XTE J1752-223) have been detected in the first year of observation. We describe the instrument, outline the technique, 
and present light curves for the seven sources.

\section{GBM AND THE EARTH OCCULTATION OBSERVATIONAL TECHNIQUE}

The Gamma ray Burst Monitor is the secondary instrument onboard the Fermi satellite \citep{Meegan2009,Wilson2009}. It consists of 12 NaI detectors 5$^{\prime\prime}$ in diameter by 0.5$^{\prime\prime}$ thick mounted on the 
corners of the spacecraft and oriented such that they view the entire sky not occulted by the Earth. GBM also contains 2 BGO detectors 5$^{\prime\prime}$ in diameter by 5$^{\prime\prime}$ thick located 
on opposite sides of the spacecraft. None of the GBM detectors have direct imaging capability.

\begin{figure}[t]
\includegraphics[trim = 10mm 0mm 15mm 146mm, clip,width = 85mm]{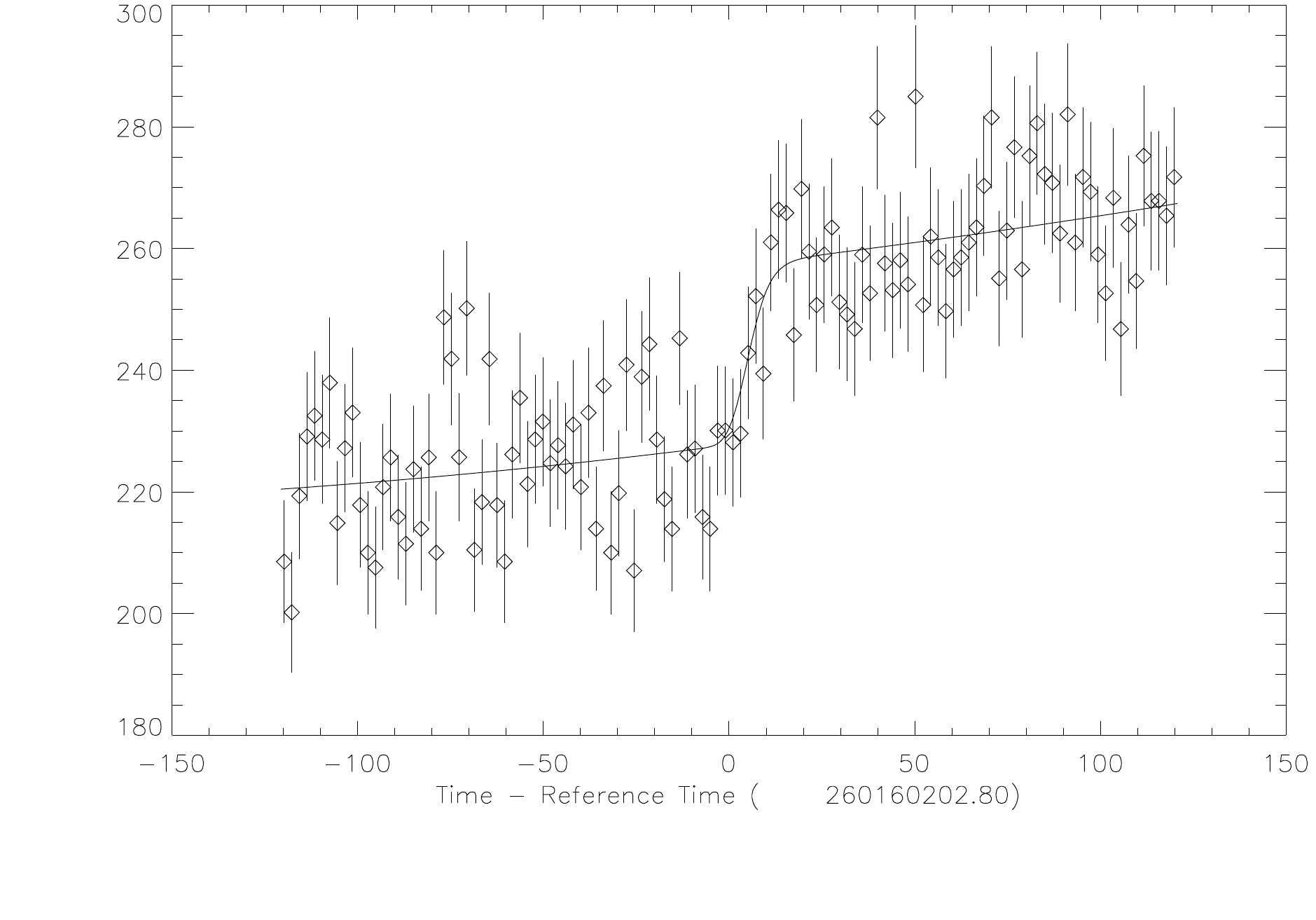}
\caption{\label{Crabstep}Single Crab occultation step in a single GBM NaI detector. Horizontal scale is in seconds centered on the occultation time. Vertical scale is in measured counts.}
\end{figure}

Known sources of gamma ray emission can be monitored with non-imaging detectors using the Earth occultation technique, as was successfully demonstrated with 
BATSE \citep{Harmon2002,Harmon2004}. When a source of gamma rays is occulted by the Earth, the count rate measured by the detector will drop, producing a step-like feature. When the 
source reappears from behind the Earth's limb, the count rate will increase, producing another step. The diameter of the Earth seen from Fermi is $\sim 140^\circ$, so roughly 
30\% of the sky is occulted by the Earth at any one time.  Coupled with the $\pm 35^\circ$ slewing of the pointing direction every orbit, this means that the entire sky is 
occulted every two orbits. With an altitude of 565 km, a period of 96 minutes, and an orbital inclination of $26.5^\circ$, individual occultation steps last for $\sim$10 seconds (Fig.\ \ref{Crabstep}).

The shape of the individual occultation steps depends on energy and occultation angle. Transmission as a function of time is modeled as $T(t) = exp[-\mu(E) A(h)]$, where $\mu(E)$ is the 
mass attenuation coefficient of gamma rays at energy $E$ in air and $A(h)$ is the air mass along the line of sight at a given altitude $h(t)$. Account is taken of the  detector response as it changes as a function of angle across the fit window. For each source, occultation times are predicted. 
Each step is fit over a 4-minute window along with a quadratic background and using an assumed spectrum to determine the detector count rate due to the source. The instrument 
response is used to convert the count rate to a flux. Up to 31 steps are possible for a given source in a day, and these steps are summed to get a single daily average flux. The GBM 
occultation sensitivity exceeds that of BATSE at energies below $\sim25$ keV and above $\sim1.5$ MeV \citep{Case2007}.

This work uses the GBM CTIME data, with its 8 broad energy channels and 0.256-second resolution, rebinned to 2-second resolution. The occultation technique relies on an input 
catalog of known sources. Currently, we are monitoring 64 sources. Of these 64 sources, 6 steady sources are detected above 100 keV with a significance of at least $5\sigma$ after $\sim490$ days 
of observations, and one transient source.

\section{RESULTS}
The results presented here are preliminary. We have not completed the fine tuning of our algorithms, though the average fluxes are not expected to change much. Future work will include 
using the GBM CSPEC data, with its finer energy binning, to examine the detailed spectra for these sources.

The measured 20 - 50 keV GBM light curves are compared to Swift's 15 - 50 keV light curves for several sources over the same time intervals in ref.~\citep{Wilson2009}, where it is seen that the results 
measured by the two instruments compare well. At energies above the upper energy limit of $\sim195$ keV of the Swift 22-month catalog \citep{Tueller2010}, however, the GBM observations provide the 
only wide-field monitor available of the low energy gamma ray sky. 

\subsection{Steady Sources}
\begin{figure}[t]
\includegraphics[width=80mm]{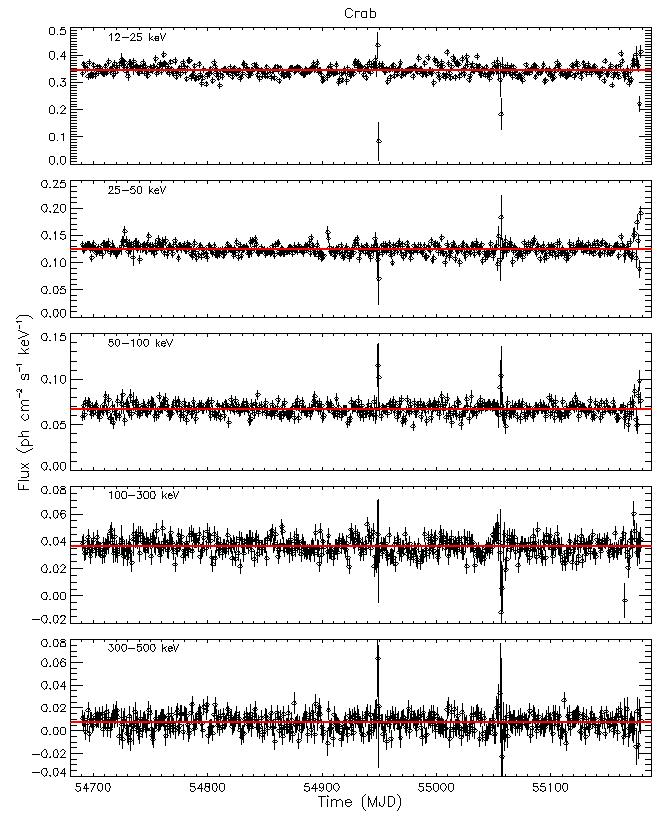}%
\caption{\label{Crab}Crab light curve. Horizontal scale is in modified Julian days over the 490 day GBM exposure period. Vertical scale is in photons/cm$^2$/sec/keV averaged over daily intervals. Horizontal lines show the average flux in each of five energy bands increasing from top to bottom}
\end{figure}

The sources Crab, Cyg X-1, Swift J1753.5-0127, 1E 1740-29, Cen A, and GRS 1915+105 are detected by GBM at energies above 100 keV. We show GBM light curves generated from the Earth occultation analysis in several energy bands with one day resolution for these six sources in Figures \ref{Crab} - 7. 

Table \ref{Flux_table} gives the 
fluxes and significances averaged over all the days from 
Aug. 12, 2008 (the beginning of science operations) to Dec. 15, 2009, approximately 490 days. 
 
The {\bf Crab} (Fig.~\ref{Crab}) spectrum in the hard x-ray/low energy gamma-ray region can be described by a broken power law, with the spectrum steepening at ~100 keV and then hardening at ~650 keV \citep{Ling2003,Jourdain2009}.  
While the GBM CTIME data do not have the spectral resolution to observe these breaks, GBM is able to see significant emission above 300 keV, consistent with the canonical hard spectrum.

\begin{table*}[t]
\begin{center}
\caption{Fluxes and Significance in High Energy Bands}
\begin{tabular}{|l|| c|c|c|| c|c|c|| c|c|c||}
\hline
    & \multicolumn{3}{c||}{50 - 100 keV} & \multicolumn{3}{c||}{100 - 300 keV} & \multicolumn{3}{c||}{300 - 500 keV} \\ \cline{2-10}
& Flux & Error & Signif. & Flux & Error & Signif. & Flux & Error & Signif. \\
& (mCrab) & (mCrab) & ($\sigma$) & (mCrab) & (mCrab) & ($\sigma$)& (mCrab) & (mCrab) & ($\sigma$) \\
\hline Crab     & 1000 & 3 & 336 & 1000 & 6 & 182 & 1000 & 47 & 21.2 \\
\hline Cen A    & 72   & 4 & 18  & 108  & 7 & 15  & 42   & 47 & 0.9  \\
\hline Cyg X-1  & 1130 & 4 & 283 & 1094 & 8 & 137 & 474  & 50 & 9.5  \\
\hline GRS 1915+105
                                & 121  & 4 & 30  & 49   & 7 & 7   & 41   & 52 & 0.8  \\
\hline 1E 1740-29  & 113  & 5 & 23  & 96   &10 & 10  & 97   & 68 & 1.4  \\
\hline SWIFT 1753.5-0127
                                & 135  & 5 & 27  & 151  & 9 & 17  & 131   & 64 & 2.0  \\
\hline XTE J1752-223
                             & 770  &16 & 48  & 622  &30 & 21  & 132  & 218 & 0.6  \\
\hline
\end{tabular}
\label{Flux_table}
\end{center}
\end{table*}

\begin{figure}
\includegraphics[width=80mm]{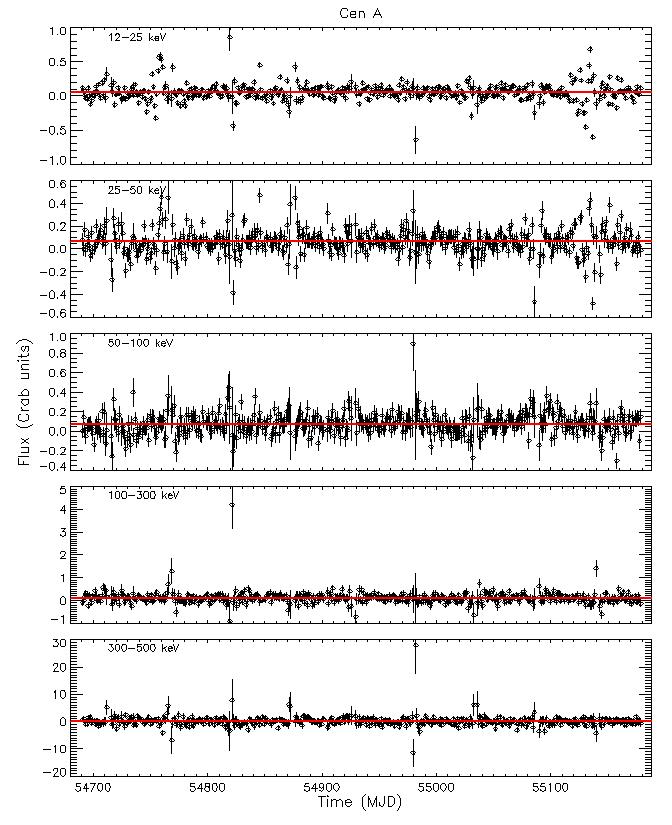}%
\caption{\label{CenA}Cen A light curve. Horizontal scale is in modified Julian days.}
\end{figure}

{\bf Cen A} (Fig.~\ref{CenA}) is a Sy 2 galaxy that is the brightest AGN in hard x-rays/low energy gamma rays.  It has a hard spectrum ($\Gamma= 1.8$) and has been observed at energies $>1$ MeV \citep{Steinle1998}.  
The GBM results are consistent with this hard spectrum, though GBM does not have the sensitivity to determine if the hard spectrum continues beyond 300 keV or if the spectrum cuts off.

\begin{figure}
\includegraphics[width=80mm]{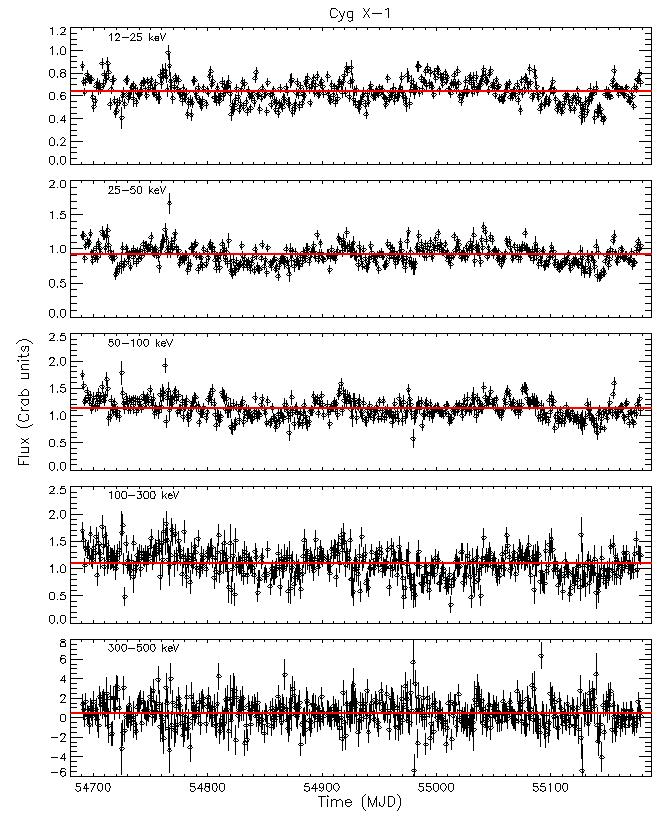}%
\caption{\label{CygX1}Cyg X-1 light curve. Horizontal scale is in modified Julian days.}
\end{figure}

{\bf Cyg X-1} (Fig.~\ref{CygX1}) is a HMXB and one of the first systems determined to contain a black hole.  It has been observed to emit significant emission above 100 keV including a power law tail extending 
out to greater than 1 MeV \citep{McConnell2000,Ling2005}.  The GBM results show significant emission above 300 keV, consistent with the power law tail observed when Cyg X-1 is in its hard state.

\begin{figure}
\includegraphics[width=80mm]{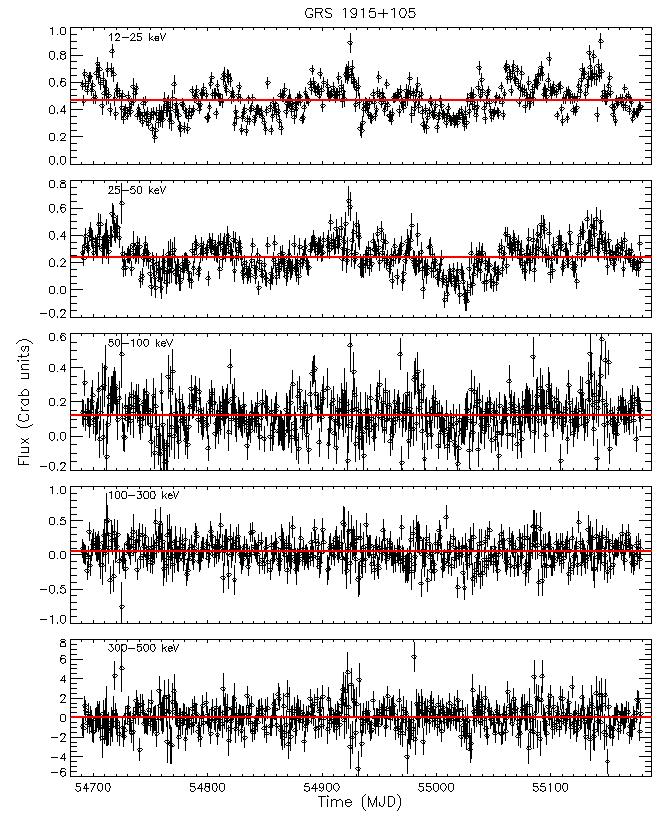}%
\caption{\label{GRS1915}GRS 1915+105 light curve. Horizontal scale is in modified Julian days.}
\end{figure}

{\bf GRS 1915+105} (Fig.~\ref{GRS1915}) is a LMXB with the compact object being a massive black hole.  Evidence for emission above 100 keV has been seen previously \citep{Case2005} with BATSE.  The GBM light curve 
integrated over ~490 days shows significant emission above 100 keV.

\begin{figure}
\includegraphics[width=80mm]{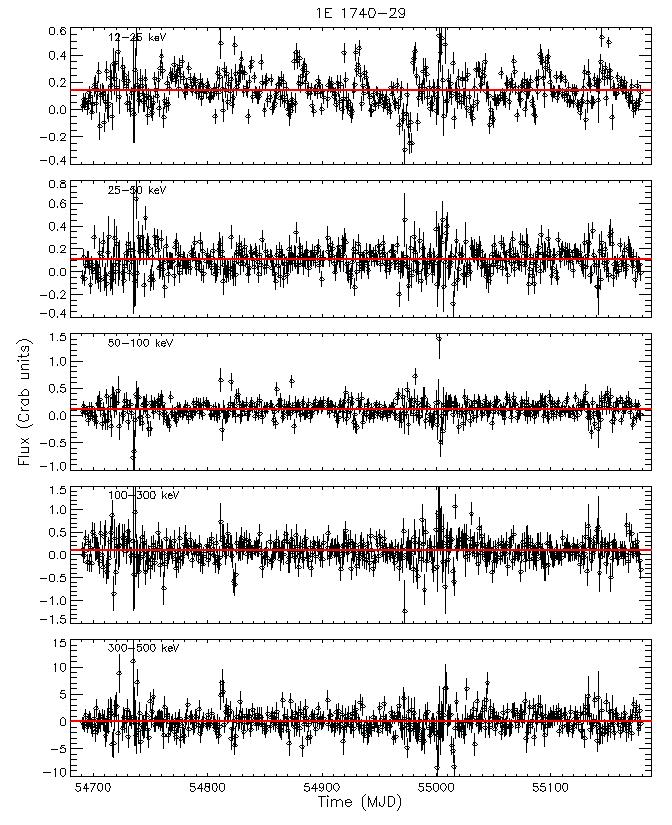}%
\caption{\label{1e1740}1E1740-29 light curve. Horizontal scale is in modified Julian days.}
\end{figure}

{\bf 1E 1740-29} (Fig.~\ref{1e1740}) is a LMXB very near the Galactic Center.  It is a microquasar, and spends most of its time in the low/hard state.  Integral observations indicate the presence 
of a power law tail above 200 keV \citep{Bouchet2009}.  The present GBM results are consistent with this high energy emission. In the future, we will use the GBM CSPEC data with their finer energy bins 
to obtain a fit to the spectrum and compare the power law index to that measured by Integral.

\begin{figure}
\includegraphics[width=80mm]{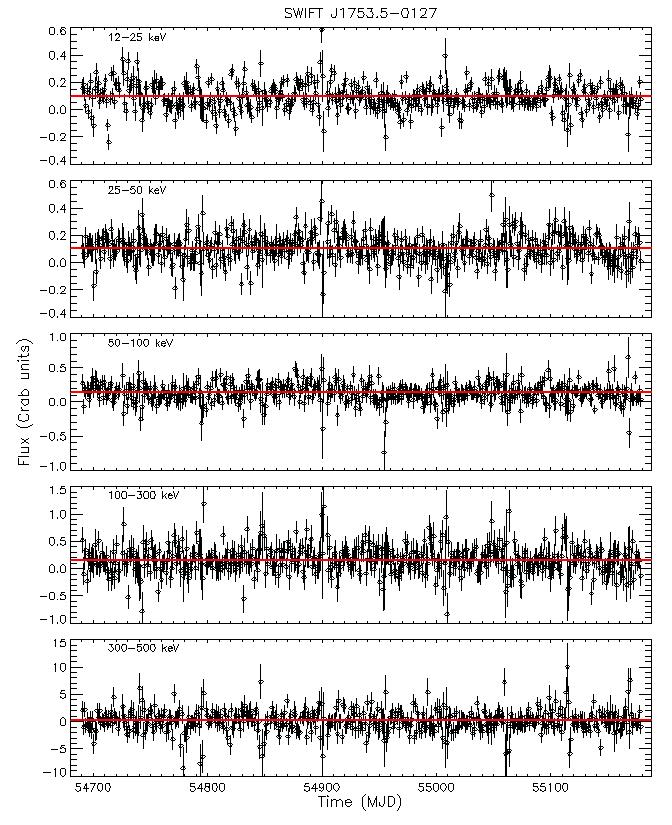}%
\caption{\label{Swift}SWIFTJ1753.5-0127 light curve. Horizontal scale is in modified Julian days.}
\end{figure}

\begin{figure}[h]
\includegraphics[width=80mm]{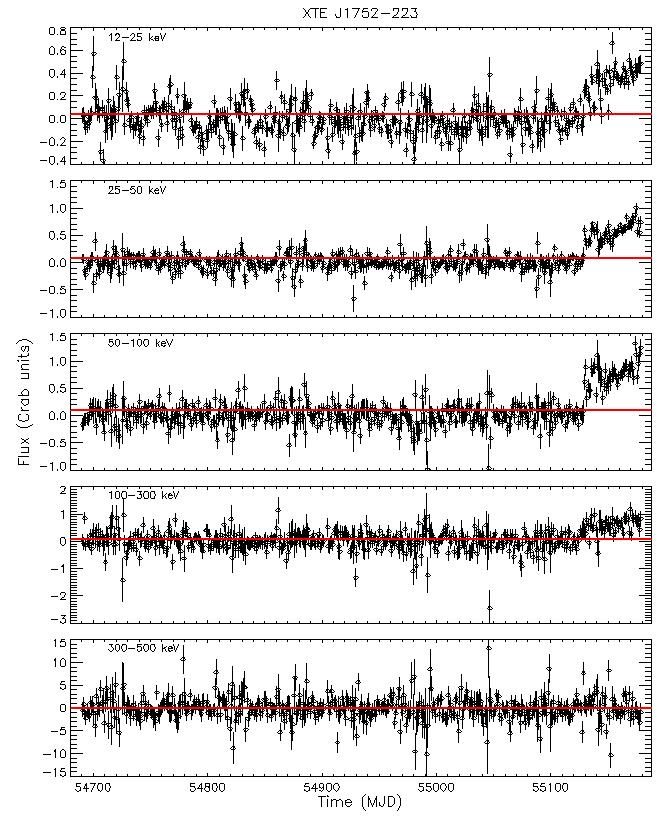}%
\caption{\label{XTEJ1752}XTEJ1752-223 light curve. Horizontal scale is in modified Julian days.}
\end{figure}

{\bf SWIFT J1753.5-0127} (Fig.~\ref{Swift}) is a LMXB with the compact object likely being a black hole.  Swift discovered this source when it observed a large flare in July of 2005.  The source did not 
return to quiescence but settled into a low intensity hard state \citep{Bel2007}.  BATSE occultation measurements from 1991-2000 showed no significant emission from this source above 25 keV \citep{Case2010}.  
The GBM results show that this source is still in a hard state, with significant emission above 100 keV.  We will continue to monitor this source while it is in the hard state, with longer 
observations potentially verifying significant emission above 300 keV.

\subsection{Transient Source}
The new transient black hole candidate {\bf XTE J1752-223}  rose from undetectable on 2009 October 24 to $511\pm 50$ mCrab (12 - 25 keV), 
$570\pm 70$ mCrab (25 - 50 keV), $970\pm 100$ mCrab (50 - 100 keV), and $330\pm 100$ mCrab (100 - 300 keV) on 2009 November  2 \citep{Wilson2009,Wilson2009b}. The light curve is 
variable, especially in the 12-25 keV band, where the flux initially rose to about 240 mCrab (2009 Oct 25-28), suddenly dropped to non-detectable on 
2009 October 29-30, then rose again during the period 2009 October 31 to November 2. As of mid December 2009, the source remains in a high intensity state. 
The light curve is shown for the period MJD 54700-55200, again with 1-day resolution, in Fig. \ref{XTEJ1752}.  The fluxes for XTE J1752-223  in Table 1 are given are for the interval of flaring activity, TJD 55130-55180. 

\begin{acknowledgments}
This work is supported by the NASA Fermi Guest Investigator program.  At LSU, additional support is provided by NASA/Louisiana Board of Regents Cooperative Agreement NNX07AT62A.
\end{acknowledgments}

\bibliography{hard_x-rays_f}


















\end{document}